\documentclass[12pt]{article}

\usepackage{latexsym,amssymb,epsfig}

\DeclareFontFamily{OT1}{rsfs10}{}
\DeclareFontShape{OT1}{rsfs10}{m}{n}{ <-> rsfs10 }{}
\DeclareMathAlphabet{\mathscript}{OT1}{rsfs10}{m}{n}

\newcommand{\eref}[1]{(\ref{#1})}

\newcommand{\tref}[1]{Table~\ref{#1}}

\newcommand{\cref}[1]{Chapter~\ref{#1}}
\newcommand{\bcenter}{\begin{center}}
\newcommand{\ecenter}{\end{center}}
\newcommand{\beq}{\begin{equation}}
\newcommand{\eeq}{\end{equation}}
\newcommand{\bea}{\begin{eqnarray}}
\newcommand{\eea}{\end{eqnarray}}
\newcommand{\bean}{\begin{eqnarray*}}
\newcommand{\eean}{\end{eqnarray*}}
\newcommand{\ba}{\begin{array}}
\newcommand{\ea}{\end{array}}
\newcommand{\ben}{\begin{enumerate}}
\newcommand{\een}{\end{enumerate}}
\newcommand{\bi}{\begin{itemize}}
\newcommand{\ei}{\end{itemize}}
\newcommand{\bd}{\begin{description}}
\newcommand{\ed}{\end{description}}
\def\fnote#1#2{\begingroup\def\thefootnote{#1}\footnote{#2}
     \addtocounter{footnote}{-1}\endgroup}

\def\IC{\mathbb{C}}

\def\IZ{\mathbb{Z}}

\def\IP{\mathbb{P}}

\def\cO{{\mathcal O}}

\def\nn{\nonumber}

\def\rk{\mbox{rk}}

\def\dirac{\slash{\! \! \! \! D}}

\def\ad{{\mathop {\rm ad}}}

\def\av{\wedge^2 V}
\def\avt{\wedge^2 \tilde{V}}

\def\to{\rightarrow}
\def\b{\beta}

\def\II{{\rlap{1} \hskip 1.6pt \hbox{1}}}

\def\ol{\overline}
\def\tx{\tilde{X}}
\def\tv{\tilde{V}}
\def\ext{{\rm Ext}}

\def\op1{{\mathcal O}_{\IP^1}}
\def\obp{{\mathcal O}_{B'}}

\def\v23{V_2 \otimes V_3}

\def\sm{SU(3)_C \times SU(2)_L \times U(1)_Y}

\begin{document}

\begin{titlepage}

\vspace{-2cm}

\title{
   \hfill{\normalsize  UPR-1087-T} \\[1em]
   {\LARGE Higgs Doublets, Split Multiplets and
	Heterotic {\Large $\sm$} Spectra
\author{Ron Donagi$^1$, Yang-Hui He$^2$, Burt A.~Ovrut$^2$, 
        and Ren\'{e} Reinbacher$^3$
        \fnote{~}{donagi@math.upenn.edu;
        yanghe, ovrut@physics.upenn.edu;
        rreinb@physics.rutgers.edu}\\[0.5cm]
   {\normalsize $^1$ 
        Department of Mathematics, University of Pennsylvania} \\
        {\normalsize Philadelphia, PA 19104--6395, USA} \\
   {\normalsize $^2$
        Department of Physics, University of Pennsylvania} \\
   {\normalsize Philadelphia, PA 19104--6396, USA} \\
   {\normalsize $^3$
        Department of Physics and Astronomy, Rutgers University}\\
   {\normalsize Piscataway, NJ 08855-0849, USA}}
}
\date{}
}

\maketitle

\begin{abstract}
A methodology for computing the massless spectrum of heterotic vacua
with Wilson lines is presented. This is applied to a specific class of
vacua with holomorphic $SU(5)$-bundles over torus-fibered Calabi-Yau
threefolds with fundamental group $\IZ_2$. These vacua lead to low
energy theories with the standard model gauge group $\sm$ and three
families of quark/leptons. The massless spectrum is computed,
including the multiplicity of Higgs doublets.
\end{abstract}

\thispagestyle{empty}

\end{titlepage}

An important goal of heterotic string theory is to demonstrate the
existence of vacua consistent with low energy particle physics
phenomenology. This has been discussed within the context of $G
(\subset E_8)$-bundles on simply connected elliptic Calabi-Yau
threefolds in \cite{ron,FMW,z2,morez2}. 
These vacua correpond to GUT theories
with gauge groups such as $SU(5)$ and $Spin(10)$. However, to
introduce Wilson lines one must extend these results to $G$-bundles on
torus-fibered Calabi-Yau threefolds $X$ with non-trivial fundamental
group. This was done in \cite{dopw}, where $G = SU(5)$ bundles on
Calabi-Yau spaces with $\pi_1(X) = \IZ_2$ were constructed. These
vacua have low energy theories with standard model gauge group $\sm$
and three families of quarks and leptons.

However, to complete the analysis of these vacua, it is essential to
compute the entire massless spectra. This was done within the GUT
context in \cite{spec}, where methods for determining the spectrum were
introduced and used in an $SU(5)$ example. An interesting result is
that the non-chiral part of the spectrum was shown to jump on isolated
subspaces of the $G$-bundle moduli space. In this paper, these methods
are extended to vacua admitting Wilson lines. We show how one computes
the massless spectrum, and give an explicit example using the $\sm$
standard model vacuum introduced in \cite{dopw}. Here, we simply
outline our results, presenting the specific details in \cite{future}.

A vacuum of heterotic string theory is determined by specifying a
Calabi-Yau threefold $X$ with fundamental group
\beq
\pi_1(X) = F
\eeq
and a stable, holomorphic vector bundle $V$ with structure group
\beq
G \subset E_8.
\eeq
Denote by
\beq
H = Z_{E_8}(G)
\eeq
the commutant of $G$ in $E_8$. $V$ is constrained to satisfy the
anomaly cancellation condition that
\beq\label{eff}
c_2(TX) - c_2(V) \quad \mbox{is effective}
\eeq
and
\beq\label{gens}
c_3(V) = 6,
\eeq
leading to three families of quark/leptons.
When $F \ne \II$, a Wilson line $W$ can be introduced. 
A Wilson line is a flat $H$-bundle. 
The bundle
\beq
V' = V \oplus W
\eeq
has structure group $G \times F$, which spontaneously breaks $E_8$ to
the gauge group
\beq
S = Z_H(F) = Z_{E_8}(G \times F).
\eeq
$X$ can be constructed as the quotient
\beq
X = \tx / F,
\eeq
where $\tx$ is a simply connected Calabi-Yau threefold and $F$ acts
freely on $\tx$. $V$ and $V'$ coincide when pulled back to $\tx$ and
are denoted by $\tv$. The structure group of $\tv$ is $G$, while
\eref{eff} and \eref{gens} become
\beq\label{cons}
c_2(\tx) - c_2(\tv) \quad \mbox{effective}, \quad
c_3(\tv) = 6 |F|
\eeq
respectively. With respect to the subgroup $G \times H \subset E_8$,
$\ad \tv$ decomposes as
\beq
\ad \tv = \bigoplus_i U_i(\tv) \otimes R_i,
\eeq
where $U_i(\tv)$ are the vector bundles associated with
the irreducible representation $U_i$ of $G$ and
$R_i$ are the corresponding representations of $H$.

As discussed in \cite{future}, the massless spectrum is identified as
\beq\label{kerD}
\ker(\dirac) = \bigoplus_{q=0,1} \bigoplus_i
\left( H^q(\tilde{X}, U_i(\tilde{V})) \otimes R_i \right)^{\rho'(F)},
\eeq
where $\dirac$ is the Dirac operator on $X$, $\rho'(F)$ specifies the
$F$ action on both $H^q(\tilde{X}, U_i(\tilde{V}))$ and $R_i$ and the
superscript indicates the invariant part of the
expression. Decomposing $R_i$ in terms of its irreducible
$F$-representations $A_j$,
\beq\label{Ri}
R_i = \bigoplus_j (A_j \otimes B_{ij}),
\eeq
expression \eref{kerD} becomes
\beq
\ker(\dirac) = \bigoplus_{q=0,1} \bigoplus_{i,j}
(H^q(\tilde{X}, U_i(\tilde{V})) \otimes A_j)^{\rho'(F)} 
	\otimes B_{ij}.
\eeq
Here, $B_{ij}$ carries a representation of the gauge group
$S$. Therefore, to compute the massless spectrum it suffices to
determine the dimension of the space of $F$-invariants in 
$H^q(\tilde{X}, U_i(\tilde{V})) \otimes A_j$.

In this paper, we choose
\beq
F = \IZ_2, \quad G = SU(5).
\eeq
Then
\beq
H = SU(5)
\eeq
and \eref{kerD} becomes 
\bea\label{spec}
\ker(\dirac) & = 
\left( H^1(\tx, U_i(\tv)) \otimes \II \right)^{\rho'(\IZ_2)}
\oplus
\left( H^0(\tx, \cO_{\tx}) \otimes 24 \right)^{\rho'(\IZ_2)}
\oplus
\left( H^1(\tx, \tv) \otimes \ol{10} \right)^{\rho'(\IZ_2)}
\nn \\
& \oplus
\left( H^1(\tx, \tv^*) \otimes 10 \right)^{\rho'(\IZ_2)}
\oplus 
\left( H^1(\tx, \avt ) \otimes 5 \right)^{\rho'(\IZ_2)}
\oplus
\left( H^1(\tx, \avt^*) \otimes \ol{5} \right)^{\rho'(\IZ_2)}.
\eea
To determine the massless spectrum, one must compute the cohomology
groups in \eref{spec}, the action of $\IZ_2$ on these groups and the
action of $\IZ_2$ on each representation $R_i$. Since the last of
these is straightforward, we discuss it first.

For $F= \IZ_2$, $W$ spontaneously breaks $H$ to the standard model
gauge group
\beq
S = \sm.
\eeq
The action of $\IZ_2$ on each representation $R_i$ of $H$ is easily
computed. For example, for $R_i = 5$ expression \eref{Ri} is
\beq
5 = 1 \otimes (3,1)_{-2} \oplus (-1) \otimes (1,2)_{3},
\eeq
where $\pm 1$ are the representations $A_j$ of $\IZ_2$ while $(a,b)_w$
are representations of $S$. 
For notational simplicity, we display $w = 3Y$.
The action of $\IZ_2$ on each
representation $R_i$ in \eref{spec}, as well as the corresponding
representations $B_{ij}$ of $S$, are listed in \tref{t:decomp}.

\begin{table}[h]
\[
\ba{||c|c|c|c|c||}\hline\hline
U_i & H^q(\tilde{X}, U_i(\tilde{V})) & R_i & \chi_{A_j} & B_{ij} \\ \hline
\hline
24 & H^1(\tilde{X}, \ad \tilde{V}) & 1 & 0 & (1,1)_0 \\ \hline
1 & H^0(\tilde{X}, \cO_{\tx}) & 24 & 0 & 
	(8,1)_0\oplus(1,3)_0\oplus(1,1)_0 \\ \hline
	& & & 1 & (3,2)_{-5}\oplus(\ol 3,2)_{5} \\ \hline
10 & H^1(\tilde{X}, \avt) & 5 & 0 & (3,1)_{-2} \\ \hline
	& & & 1 & (1,2)_{3} \\ \hline
\ol{10} & H^1(\tilde{X}, \avt^*) & \ol 5 & 0 & (\ol 3,1)_{2} \\ \hline
	& & & 1 & (1,2)_{-3} \\ \hline
5 & H^1(\tilde{X}, \tv) & \ol{10} & 0 & (3,1)_{4} \oplus (1,1)_{-6} \\ \hline
	& & & 1 & (\ol 3,2)_{-1} \\ \hline
\ol 5 & H^1(\tilde{X}, \tv^*) & 
	10 & 0 & (\ol 3,1)_{-4} \oplus (1,1)_{6} \\ \hline
	& & & 1 & (3,2)_{1} \\ \hline \hline
\ea
\]
\caption{The decomposition of $H^q(X, \ad V')$ where $G=SU(5)$
and $F=\IZ_2$.
The $\chi_{A_j}$ are the characters of the $\IZ_2$ action on $R_i$.
The $a,b$ in $(a,b)_w$ are the representations of $SU(3)_C$ and
$SU(2)_L$ respectively, whereas $w = 3Y$.}\label{t:decomp}
\end{table}

To compute the cohomology groups in \eref{spec}, we must construct
$\tx$ and $\tv$. Choose $\tx$ to be the fiber product
\beq\label{tx}
\tx = B \times_{\IP^1} B'
\eeq
of two $d\IP_9$ surfaces $B$ and $B'$. $\tx$ is elliptically fibered
over both surfaces with the projections
\beq
\pi' : \tx \to B, \quad
\pi : \tx \to B'.
\eeq
$B$ and $B'$ are themselves elliptically fibered over $\IP^1$ with the
maps
\beq\label{betaB}
\b : B \to \IP^1, \quad
\b' : B' \to \IP^1.
\eeq
A $\IZ_2$ action $\tau$ on $\tx$ can be obtained as the lift
\beq\label{tau}
\tau = \tau_B \times_{\IP^1} \tau_{B'}
\eeq
of two involutions $\tau_B$ and $\tau_{B'}$ on $B$ and $B'$
respectively. It is sufficient to know that $\tau_B$ acts on $\IP^1$
as
\beq
t_0 \to t_0, \quad
t_1 \to -t_1,
\eeq
where $t_0, t_1$ are projective coordinates. This action has two fixed
points, $p_0$ and $p_\infty$. The fiber $f_0 = \b^{-1}(p_0)$ is acted
on freely by $\tau_B$, whereas $f_\infty = \b^{-1}(p_\infty)$ has four
fixed points. The $\tau_{B'}$ action on $B'$ has similar
properties. In order for $\tau$ in \eref{tau} to act freely on $\tx$,
one must ``twist'' the two $\IP^1$ lines in \eref{betaB} when
identifying them in \eref{tx}. This twist sets $p_0' = p_\infty$ and
$p'_\infty = p_0$.

Stable, holomorphic vector bundles $\tv$ on $\tx$ with structure group
$G = SU(5)$ can be constructed as the extension
\beq\label{v23}
0 \to V_2 \to \tv \to V_3 \to 0
\eeq
of two vector bundles
\beq\label{defVi}
V_i = \pi'^*W_i \otimes \pi^* L_i
\eeq
with rk$V_i = i$ for $i=2,3$. $W_2$ and $W_3$ are vector bundles on
$B$ with rank 2 and 3 respectively, while $L_2$ and $L_3$ are line
bundles on $B'$. 
We need to lift the $\IZ_2$ action on $\tx$ to an action on $V_2$, $V_3$
and $\tv$.
For $\tv$ to be $\IZ_2$ invariant, it is necessary to restrict both
$V_2$ and $V_3$ to be invariant. This is done by choosing $W_i$ and
$L_i$ to be $\tau_B$ and $\tau_{B'}$ invariant respectively.
There are many line bundles $L_i$ that are invariant under $\tau_{B'}$.
However, we now impose the remaining constraint that $\tv$ satisfy
\eref{cons} with $|F|=2$.
This restricts the allowed line bundles to be
\beq\label{L23}
L_2 = \obp(3r'), \quad 
L_3 = \obp(-2r'), 
\eeq
where $r'$ is a specific divisor of $B'$ with $\deg(r')=2$ when
restricted to a fiber.
By construction, $\tv$ corresponds to an extension class
\beq
[\tv] \in \ext^1_{\tx}(V_3, V_2).
\eeq
$\ext^1_{\tx}(V_3, V_2)$ is the direct sum of two subspaces which are
invariant and anti-invariant under the action of $\tau$.
$\tv$ will be invariant if $[\tv]$ lies in the invariant subspace.

We can now construct the cohomology groups in \eref{spec}. However,
one group, $H^1(\tx, \ad \tv)$, corresponding to vector bundle moduli,
requires techniques beyond those developed in this paper and will not
be discussed.
Let us consider $H^0(\tx, \cO_{\tx})$.
Since $\cO_{\tx}$ is the trivial bundle, it follows that
\beq
H^0(\tx, \cO_{\tx}) \simeq \IC.
\eeq
Note that since $\cO_{\tx}$ is independent of $\tv$, $\IZ_2$ acts
trivially on $H^0(\tx, \cO_{\tx})$.

Next, we determine $H^1(\tx, \tv)$.
From the long exact sequence associated with \eref{v23}, we find that
\beq
H^1(\tx, \tv) \simeq H^1(\tx, V_2).
\eeq
Using \eref{defVi} and pushing this down from $\tx$ to $\IP^1$ gives
\beq
H^1(\tx, V_2) \simeq H^0(\IP^1, R^1 \b_* W_2 \otimes \b'_* L_2).
\eeq
We find that $R^1\b_*W_2 \simeq \cO_{p_\infty}$. 
From \eref{L23} it follows that $\b'_* L_2$ has degree 6 along $f'_0$.
We conclude that
\beq\label{h1V}
H^1(\tx, \tv) \simeq \IC \otimes \IC^6 = \IC^6.
\eeq
Now consider $H^1(\tx, \tv^*)$.
This can be determined from \eref{h1V} using the Atiyah-Singer index
theorem which, together with Serre duality, gives
\beq
h^1(\tx, \tv^*) = 6 + h^1(\tx, \tv),
\eeq 
where we have used \eref{cons} with $|F|=2$.
This and \eref{h1V} then imply
\beq\label{h1Vd}
H^1(\tx, \tv^*) \simeq \IC^{12}.
\eeq

We now turn to the computation of $H^1(\tx, \avt)$.
One can show that $H^1(\tx, \avt)$ lies in the exact sequence
\beq\label{seqavt}
0 \to H^1(\tx,\av_2) \to 
\fbox{$H^1(\tx, \avt)$}  \to  H^1(\tx, V_2\otimes V_3)  
\stackrel{M^T}{\longrightarrow}
H^2(\tx,\av_2) \to \ldots
\eeq
To continue, we must compute the terms $H^i(\tx,\av_2)$, $i=2,3$ and
$H^1(\tx, V_2\otimes V_3)$, as well as the linear map $M^T$.
Pushing $H^i(\tx,\av_2)$, $i=2,3$ down to $\IP^1$, we find
\beq
H^1(\tx,\av_2) \simeq \bigoplus^5 H^0(\IP^1, \op1)^*
\eeq
and
\beq
H^2(\tx,\av_2) \simeq \bigoplus^7 H^0(\IP^1, \op1)^* \oplus
	\bigoplus^5 H^0(\IP^1, \op1(1))^*.
\eeq
It follows that
\beq\label{h12av2}
H^1(\tx,\av_2) \simeq \IC^5, \quad
H^2(\tx,\av_2) \simeq \IC^{17}.
\eeq
Calculating $H^1(\tx, V_2\otimes V_3)$ is more difficult. Using
\eref{defVi} and pushing down to $\IP^1$, we find
\beq\label{H1v23}
H^1(\tx, V_2\otimes V_3) \simeq 
H^0(\IP^1, R^1 \b_*(W_2 \otimes W_3) \otimes \b'_*(L_2 \otimes L_3)).
\eeq
Here, we simply state that $ R^1 \b_*(W_2 \otimes W_3)$ is a sheaf
supported at each of 12 points $p_r \in \IP^1$,
$r=1,\ldots,12$ and at $p_\infty$.
Specifically,
\beq\label{R1bw23}
R^1 \b_*(W_2 \otimes W_3) \simeq 
\bigoplus_{r=1}^{12} \cO_{p_r} \oplus \bigoplus^3 \cO_{p_\infty}.
\eeq
Furthermore, it follows from \eref{L23} that $\b'_*(L_2 \otimes L_3)$
is a rank two vector bundle on $\IP^1$.
Combining these results, \eref{H1v23} becomes
\beq\label{h1v23}
H^1(\tx, V_2\otimes V_3) \simeq \IC^{15} \otimes \IC^2 = \IC^{30}.
\eeq
Finally, we must know the rank of $M^T$. It follows from \eref{h12av2}
and \eref{h1v23} that $M^T$ is a $30 \times 17$ matrix.
In addition, one can show it depends on 150 vector bundle
moduli.
At a generic point in moduli space, we find that
\beq\label{rkM}
\rk(M^T) = 17.
\eeq
Putting \eref{h12av2}, \eref{h1v23} and \eref{rkM} into \eref{seqavt},
we conclude that
\beq\label{h1avt}
H^1(\tx, \avt) \simeq \IC^{18}.
\eeq
Finally, we need to compute $H^1(\tx, \avt^*)$. Again, this is easily
determined using the Atiyah-Singer index theorem.
In this context, we find
\beq
h^1(\tx, \avt^*)= 6 + h^1(\tx, \avt).
\eeq
Combining this with \eref{h1avt} yields
\beq
H^1(\tx, \avt^*) \simeq \IC^{24}.
\eeq

Having computed all the cohomology groups in \eref{spec}, we now
determine the explicit action of $\IZ_2$ on each of them.
Let us begin with $H^1(\tx,\tv)$, which was given in \eref{h1V}.
To begin with, consider the second factor, $\IC^6$.
This can be shown to be parametrized by the
polynomials
\beq\label{basisC6}
\{x_0^{3-i} x_1^{i},~~ y x_0^{1-j} x_1^{j} \},
\eeq
where $i=0,\ldots,3$ and $j=0,1$.
Here, $x_0, x_1$ and $y$ are sections of specific bundles on the base
$\IP^1$, which transform as
\beq
x_0 \to x_0, \quad
x_1 \to -x_1, \quad
y \to y
\eeq
under $\tau_{B'}$.
Applying these transformations to \eref{basisC6}, we see that $\IC^6$
decomposes as $\IC_{(+)}^{3} \oplus \IC_{(-)}^{3}$ under the action of
$\IZ_2$.
Since this is evenly split between $+$ and $-$, the $\IZ_2$ action on the
first factor $\IC$ in \eref{h1V} is irrelevant.
We conclude that
\beq\label{h1V-res}
H^1(\tx,\tv) \simeq \IC_{(+)}^{3} \oplus \IC_{(-)}^{3}.
\eeq

We now compute the $\IZ_2$ action on $H^1(\tx,\tv^*)$ in \eref{h1Vd}
using the Atiyah-Singer index theorem.
First, consider the index theorem for $V$ on $X= \tx / \IZ_2$.
Using \eref{cons} with $|F|=2$, the fact that
$H^q(\tx,\tv)_{(+)}=H^q(X,V)$ for any $q$ and Serre duality, we find
\beq\label{indexv+}
h^1(\tx, \tv^*)_{(+)} = 3 + h^1(\tx, \tv)_{(+)}.
\eeq
Using \eref{cons}, Serre duality and \eref{indexv+}, the index theorem
for $\tv$ on $\tx$ becomes
\beq\label{indexv-}
h^1(\tx, \tv^*)_{(-)} = 3 + h^1(\tx, \tv)_{(-)}.
\eeq
It then follows from \eref{h1V-res}, \eref{indexv+} and \eref{indexv-}
that
\beq\label{h1Vd-res}
H^1(\tx, \tv^*) \simeq \IC_{(+)}^{6} \oplus \IC_{(-)}^{6}
\eeq
under the action of $\IZ_2$.

Now consider $H^1(\tx, \avt)$ in \eref{h1avt}. 
It follows from
\eref{seqavt} that to find the $\IZ_2$ action on $H^1(\tx, \avt)$, one
must determine its action on $H^i(\tx, \av_2)$, $i=1,2$ in
\eref{h12av2}, $H^1(\tx, V_2\otimes V_3)$ in \eref{h1v23} and on the
map $M^T$ satisfying \eref{rkM}.
Since the decomposition of each of these cohomology groups under
$\IZ_2$ is computed using methods similar to those leading to
\eref{h1V-res}, we simply state the results.
We find
\beq\label{h1av2+-}
H^1(\tx, \av_2) \simeq \IC_{(+)}^{3} \oplus \IC_{(-)}^{2}, \quad
H^2(\tx, \av_2) \simeq \IC_{(+)}^{9} \oplus \IC_{(-)}^{8}
\eeq
and
\beq\label{h1v23+-}
H^1(\tx, V_2\otimes V_3) \simeq 
	\IC_{(+)}^{15} \oplus \IC_{(-)}^{15}.
\eeq
Furthermore, one can show that $M^T$ can be taken to be invariant
under $\IZ_2$, corresponding to choosing $[\tv]$ to be
in $\ext^1_{\tx}(V_3, V_2)_{(+)}$.
Then, it follows from \eref{h1av2+-} and \eref{h1v23+-} that
\beq\label{M+:+-}
(\ker M^T)_{(+)} = \IC_{(+)}^{6},  \quad
(\ker M^T)_{(-)} = \IC_{(-)}^{7}.  
\eeq
Putting \eref{h1av2+-} and \eref{M+:+-} into the exact sequence
\eref{seqavt}, we conclude that
\beq\label{h1avt+-}
H^1(\tx, \avt) \simeq \IC^{9}_{(+)} \oplus \IC^{9}_{(-)}.
\eeq

Finally, we can compute the $\IZ_2$ action on $H^1(\tx, \avt^*)$ using
the Atiyah-Singer index theorem. This computation is very similar to
that leading to \eref{h1Vd-res}, so we will simply state the result.
We find that 
\beq\label{h1avtd+}
H^1(\tx, \avt^*) \simeq \IC^{12}_{(+)} \oplus \IC^{12}_{(-)}.
\eeq

\begin{table}[h]
\[
\ba{||c|c|c|c||}\hline\hline
R_i & (\chi_{H^q},A_j) &  
(H^q(\tilde{X}, U_i(\tilde{V})) \otimes A_j)^{\rho'(F)} & B_{ij}
\\ \hline \hline
1 & & & 
	\\ \hline
24 & (0,0)  & \IC_{(+)}^{1} & 
	(8,1)_0 \oplus (1,3)_0 \oplus (1,1)_0
	\\ \hline
5 & (0,0) & \IC_{(+)}^{9} & (3,1)_{-2}
	\\ \hline
& (1,1) & \IC_{(-)}^{9} & (1,2)_{3}
	\\ \hline
\ol 5 & (0,0) & \IC_{(+)}^{12} & (\ol 3,1)_{2}
	\\ \hline
& (1,1) & \IC_{(-)}^{12} & (1, 2)_{-3}
	\\ \hline
\ol{10} & (0,0) & \IC_{(+)}^{3} & (3,1)_{4} \oplus (1,1)_{-6}
	\\ \hline
& (1,1) & \IC_{(-)}^{3} & (\ol 3, 2)_{-1}
	\\ \hline
10 & (0,0) & \IC_{(+)}^{6} & (\ol 3,1)_{-4} \oplus (1,1)_{6}
	\\ \hline
& (1,1) & \IC_{(-)}^{6} & (3, 2)_{1}
	\\ \hline 
\hline
\ea
\]
\caption{The particle spectrum of the low-energy $\sm$ theory.
The $\chi_{H^q}$ are the
characters of the $\IZ_2$ representations on 
$H^q(\tilde{X}, U_i(\tilde{V}))$.
The $U(1)$ charges listed are $w = 3Y$.
}
\label{t:res}
\end{table}

We now possess all of the ingredients necessary to compute the
massless spectrum.
Combining \eref{h1V-res}, \eref{h1Vd-res} and
\eref{h1avt+-}-\eref{h1avtd+}  
with the results in \tref{t:decomp}, one can determine
the $\rho'(\IZ_2)$ invariant subspace for each cohomology group in
\eref{spec}.
The associated multiplets descend to $X = \tx / \IZ_2$ to form the
$\sm$ particle physics spectrum.
The results are tabulated in \tref{t:res}.

To begin with, the spectrum contains one copy of vector supermultiplets
transforming under $\sm$ as
\beq
(8,1)_0 \oplus  (1,3)_0 \oplus (1,1)_0.
\eeq
Furthermore, it contains three families of quarks and lepton
superfields, each family transforming as
\beq
(3,2)_{1}, \quad (\ol{3},1)_{-4}, \quad (\ol{3},1)_{2}
\eeq
and
\beq
(1,2)_{-3}, \quad (1,1)_{6}
\eeq
respectively.
However, there are additional chiral superfields in the spectrum.
It follows from \tref{t:res} that these occur as conjugate pairs of
the $\sm$ representations
\beq\label{5}
(3,1)_{-2}, \quad (1,2)_{3}
\eeq
and
\beq\label{10bar}
(3,1)_{4} \oplus (1,1)_{-6}, \quad (\ol{3},2)_{-1}.
\eeq
These multiplets arise as $\IZ_2$ invariants in the 5 and $\ol{10}$
representations of $H = SU(5)$.
The spectrum has 
\beq
n_{(3,1)_{-2}} = 9, \quad
n_{(1,2)_{3}} = 9
\eeq
and
\beq
n_{(3,1)_{4} \oplus (1,1)_{-6}} = 3, \quad
n_{(\ol 3,2)_{-1}} = 3
\eeq
copies of \eref{5} and \eref{10bar} respectively.
The multiplicity $n_{(1,2)_{3}}$ corresponds to the number of Higgs
doublet conjugate pairs in the low energy spectrum.
The remaining multiplets in \eref{5} and \eref{10bar} are exotic.
Clearly, the number of Higgs doublets and the exotic multiplets is
not consistent with phenomenology.
However, we emphasize that these results were computed within a
specific context, which is but a small subset of
the possible standard model heterotic vacua.
These generalized vacua and their spectra will be presented in future
publications.

\paragraph{Acknowledgements}
We are grateful to Volker Braun and Tony Pantev for enlightening
discussions.
R.~D.~would like to acknowledge conversations with Jacques Distler.
This Research was supported in part by
the Department of Physics and the Maths/Physics Research Group
at the University of Pennsylvania
under cooperative research agreement DE-FG02-95ER40893
with the U.~S.~Department of Energy and an NSF Focused Research Grant
DMS0139799 for ``The Geometry of Superstrings.'' R.~D.~further
acknowledges an NSF grant DMS 0104354.
R.~R.~is also supported
by the Department of Physics and Astronomy of Rutgers University under
grant DOE-DE-FG02-96ER40959.

\bibliographystyle{JHEP}

\end{document}